\documentstyle[11pt,psfig]{article}

\newcommand{\be}{\begin{equation}}
\newcommand{\ee}{\end{equation}}
\newcommand{\beq}{\begin{eqnarray}}
\newcommand{\eeq}{\end{eqnarray}}
\newcommand{\bed}{\begin{displaymath}}
\newcommand{\eed}{\end{displaymath}}
\begin{document}

\thispagestyle{empty}

\begin{center}

\vspace{1.0in}

{\Large\bf{Long-wavelength approximation for string cosmology
with barotropic perfect fluid }}

\vspace{0.8in}

Piret Kuusk$^{1}$\\
Institute of Physics, University of Tartu,
Riia 142, Tartu 51014, Estonia\\
Margus Saal$^{2}$\\
University of Tartu, T\"ahe 4, Tartu 51010, Estonia

\vspace{0.3in}


\end{center}

\vspace{0.5in}

\begin{abstract}

The field equations derived from the low energy string
effective action with a matter tensor describing a perfect fluid 
with a barotropic equation of state 
are solved iteratively using the 
long-wavelength approximation, i.e.  
the field equations are expanded by the number of spatial gradients. 
In the zero order, a quasi-isotropic solution is presented and 
compared with the general solution of the pure dilaton gravity.
Possible cosmological models are analyzed from the point of view of
the pre-big bang scenario. 
The second order solutions are found 
and their growing and decaying parts are studied.
\end{abstract}

\vspace{1.0in}

$^1$ Electronic address: piret@fi.tartu.ee

$^2$ Electronic address: margus@hexagon.fi.tartu.ee

\newpage

\section {Introduction} \label{uks}

The long-wavelength iteration scheme for investigating
the early stages of cosmological models was first introduced 
by Lifschitz and Khalatnikov \cite{lk}
in the case of the Einstein equations with pure radiation as a source term.
An improved scheme was presented by Tomita \cite{Tomita}
where its main assumption was taken to be 
the smallness of spatial variations of metric 
$\partial_{l} \gamma_{ij}$ 
compared with time variations 
$\partial_{0} \gamma_{ij}$. 
Then Comer, Deruelle, Langlois, and Parry 
\cite{cdlp}  applied it to the Einstein equations with a perfect 
fluid matter (with a barotropic equation of state, cf also \cite{lks}) 
and  in the case  of a scalar field with a potential. 
The next step was done by Tomita and Deruelle \cite{td} who 
examined the case of two fluids (an inflationary and a standard fluid).
Finally, the scalar-tensor theory with a perfect fluid has been 
discussed by Comer, Deruelle, and Langlois  \cite{cdl}. 
The long-wavelength approximation for the Hamilton-Jacobi  
formalism (the gradient expansion for the generating functional) 
was developed by Salopek, Stewart, and Parry \cite{ssp}. 

In the following, we solve iteratively  field equations derived from 
the low energy string effective action with a perfect fluid matter
with additional requirement that the matter obeys a barotropic equation of 
state.  
In general, the approximation scheme can be applied 
in the case of anisotropic and 
inhomogeneous spacetimes.  In the present paper we  ignore 
the local anisotropy, although this assumption 
may be incorrect near singularity \cite{dl}.   
We find a quasi-isotropic solution for 
the field equations in the zero
approximation (neglecting all spatial gradients) 
which include both
post-big bang and pre-big bang  
\cite{homepage, lidsey, sfd, pbb} cosmological models. 
It has been demonstrated \cite{bv, kmo} that a 
smooth transition from a pre-big bang branch to a post-big bang branch 
(the graceful exit problem, see also \cite{ge, WdW, ellis})  
is impossible in this approximation,
even if we incorporate some dilaton potential, or cosmological constant.
We analyze the  nature of discontinuities 
which are contained in cosmological models based on 
a quasi-isotropic solution
with a barotropic perfect fluid as a matter source (cf \cite{is}).
  
Recently, Khoury, Ovrut, Seiberg, Steinhardt, and Turok \cite{kosst}
presented a new cosmological scenario 
where the Universe is initially contracting towards a big 
crunch and then makes a transition through a singularity
to the post-big bang Universe. 
The  model is a part of the M-theory motivated ekpyrotic
scenario developed by Khoury, Ovrut,  Steinhardt,  and Turok 
\cite{ekpyrotic} and occurs in a recent proposal which introduces 
the cyclic nature of the ekpyrotic Universe \cite{cyclic}.   
The same considerations
can be applied to the reversal problem in the  pre-big bang scenario. 

The paper is organized as follows. 
In the next section,  the string frame field equations 
in a synchronous coordinate system are given. 
In the third section, solutions of the field equations
in the zero approximation (taking spatial gradient terms to vanish) 
in the case of pure dilaton gravity and 
in the case of dilaton gravity with a perfect fluid matter 
 obeying a barotropic equation of state are presented.
Specific interpretations of solutions are considered and 
problems of cosmological model building are studied. 
In the fourth section, the second order solutions in the long-wavelength 
approximation are found. It is analyzed how the equation of state for 
matter influences the evolution of inhomogeneities.
The fifth section is a  summary.

\section {Field Equations} \label{kaks}

Upon compactification the low energy action 
can be written  in the lowest order in the inverse string tension 
and coupling and using  the string (Jordan) 
frame as
\beq \label{3+1}
    I_{eff} &=& \frac{1}{2\lambda_{s}^{2}} \int d^{4}x \sqrt{-g} e^{-\phi} 
    [~^{4}{\mathcal{R}} + g^{\mu\nu} \partial_\mu \phi \partial_\nu \phi - 
    \frac{1}{12}H_{\mu\nu\rho}H^{\mu\nu\rho} + V(\phi)] 
\nonumber\\[2ex]
           &-& \int d^{4}x \sqrt{-g}[~L_{m} - \Lambda] .
\eeq
Here $^{4}{\mathcal{R}}$ is the curvature scalar of the metric $g_{\mu\nu}$,
$\phi$ is the dilaton field determining the strength of the gravitational
coupling through $g_{s}^{2} \sim e^{\phi}$, 
$H_{\mu\nu\rho} $ is an antisymmetric field strength, 
$\lambda_{s}$ is the 
fundamental string length scale, $L_{m}$ is the matter lagrangian,
and $\Lambda$ is the cosmological constant. 
We assume that there is no direct coupling between the dilaton and
matter fields.

The energy-momentum tensor for a perfect fluid matter is 
\beq
T_{\mu\nu} = (\rho + p) u_{\mu} u_{\nu} + p g_{\mu\nu} \label{pfluid}
\eeq 
and we assume that the energy density and pressure are related by
a barotropic equation of state
\beq
p = (\Gamma - 1) \rho , ~  0 \leq \Gamma \leq 2. \label{barotropic}
\eeq
Here $\Gamma = {\frac{4}{3}}$  corresponds to a pure radiation,  
$\Gamma = 1$  to a pressureless dust, $\Gamma = {\frac{2}{3}}$ to an
inflationary fluid,  $\Gamma = 0$ to a phenomenological 
matter sector cosmological constant, and $\Gamma = 2$ to a stiff fluid. 

Let us introduce a synchronous gauge
\beq
 ds^{2} = -dt^{2} + \gamma_{ij}(t,x) dx^{i} dx^{j}  
\eeq 
and  additional simplifying assumptions of
vanishing three-form field,   $H_{\mu\nu\rho}=0$, dilaton potential, 
$V(\phi) =0$, and cosmological constant, $\Lambda=0$.
Now the field equations following from the action functional (\ref{3+1}) 
with a barotropic perfect fluid (\ref{pfluid}),
(\ref{barotropic}) as the matter source can be written 
\beq 
{\mathcal{R}}_{0}^{0} \equiv {\frac{1}{2}} {\dot K} + {\frac{1}{4}} 
                          K_{j}^{i} K_{j}^{i} &=&
{\frac{1}{4}} e^{\phi} \rho (2- 3 \Gamma - 2 \Gamma u_{s}u^{s})   
+ {\frac{1}{4}} K \dot{\phi} 
\nonumber\\[2ex]
&+& {\frac{1}{2}} ( 3 \ddot{\phi} - \dot\phi^{2} + 
\partial_{r} \phi \partial^{r} \phi - \nabla_{r}\nabla^{r} \phi) \,, 
\label{11}
\\[2ex] \label{fe7}
{\mathcal{R}}_{i}^{0} \equiv {\frac{1}{2}} (\nabla_{i} K - \nabla_{j}K_{i}^{j}) &=& 
{\frac{1}{2}} e^{\phi} \rho \Gamma u_{i} \sqrt{1+ u_{s}u^{s}}
-{\frac{1}{2}} K_{i}^{j} \nabla_{j}\phi + \nabla_{i}\dot\phi \,,
\label{12}
\\[2ex] 
{\mathcal{R}}_{i}^{j} \equiv R_{i}^{j} + {\frac{1}{2}} \dot K_{i}^{j} + 
                          {\frac{1}{4}} K K_{i}^{j} &=&
{\frac{1}{4}} e^{\phi} \rho [(2 - \Gamma) \delta_{i}^{j}  
+ 2 \Gamma u_{i}u^{j} ] 
+ {\frac{1}{2}} K_{i}^{j} \dot{\phi} \nonumber
\eeq
\beq
+ {\frac{1}{4}} K \dot{\phi} \delta_{i}^{j} 
+ {\frac{1}{2}} \delta_{i}^{j} (\ddot{\phi} - \dot\phi^{2} 
+ \partial_{r} \phi \partial^{r} \phi - \nabla_{r} \nabla^{r} \phi)
- \nabla_{i} \nabla^{j} \phi \,,
\label{13}
\eeq
\beq
\ddot \phi + \frac{1}{2} K \dot \phi - \dot \phi^{2} 
 - \nabla_{r} \nabla^{r} \phi + \partial_{r} \phi \partial^{r} \phi 
= \frac{1}{2} e^{\phi} [(3 \Gamma - 4) \rho ]\,.
\label{14}
\eeq
Here $K_{ij} = {\dot\gamma}_{ij}$, $K = K^{i}_{i} = 
\gamma^{ij} {\dot \gamma}_{ij}$ and $R_{i}^{j}$ is the Ricci tensor of
the three-metric $\gamma_{ij}$.
In addition, we have the usual conservation law for the matter
density $\nabla_{\mu} T^{\mu}_{\nu} = 0$. 


\section {Zero Order Solutions} \label{kolm}

In the zero order approximation we ignore  spatial gradients and 
local anisotropy in field equations (\ref{11})--(\ref{14}). 
A solution for the metric is taken to be in a quasi-isotropic form
\beq \label{qi}
 \gamma_{ij}(t, x^{k}) = a^{2}(t) \ h_{ij}(x^{k}) , 
\eeq
where $h_{ij}(x^k)$ is a time independent seed metric and $a(t)$  
is a scale factor of an isotropic and homogeneous Universe.  
As demonstrated in \cite{Tomita, cdlp}, the quasi-isotropic solution
(\ref{qi}) can be regarded as an attractor of general anisotropic
solutions.
We will consider only a particular class of solutions with 
a power law evolution of the scale factor and a logarithmically evolving 
dilaton 
\beq \label{sol1}
  a = a_{0}\ \tau^{\alpha(\Gamma)} , \qquad \phi = \phi_{0} - 
          \beta(\Gamma) \ln \tau ,
\eeq
where $\tau$ is a time parameter.
Solutions of this form can be found only in the case 
of the pure dilaton gravity with a vanishing potential, $V(\phi) =0$,
and with a perfect barotropic fluid
matter (\ref{pfluid}), (\ref{barotropic}) as a source.
Inclusion of the cosmological constant $\Lambda$ introduces 
a hyperbolic nature of  solutions and we shall not consider this case here.
Solutions  (\ref{sol1}) inevitably contain  
singular points ($a \rightarrow 0$, $g_s^2 \sim e^{\phi} \rightarrow \infty$) 
and it is not possible to avoid them.

\subsection{Pure Dilaton Gravity}

The field equations for the Hubble parameter 
$H(t) \equiv \frac{\dot a(t)}{a(t)}$ and the dilaton $\phi (t)$ 
can be written as follows
\beq
\dot{H} + 3H^2 - H \dot{\phi} = 0 ,\\
6 H^{2} - 6 H \dot{\phi} + \dot{\phi}^{2} = 0 ,\\
\ddot{\phi} + 3 H \dot{\phi} - {\phi}^{2} = 0 .
\eeq
The system is not overdetermined, since only two equations are independent. 

The most general isotropic solution reads
\be \label{dg2}
a = a_{0} (A t + t_{0})^{\frac{B}{\sqrt{3}}}, 
\ee
\be \label{dg1}
\phi = \phi_{0} + (B\sqrt{3} - 1) \ln (A t + t_{0}) .
\ee
Here $B=\pm 1$ denote two branches of  solutions. 
The range of variation of the time coordinate $t$ must be
chosen such that $At + t_0 \geq 0$,  so 
the time inversion amounts
to a change of  constant $A \rightarrow -A$. 
The dilaton gravity analogue
of a post-big bang model corresponds to $A > 0, B=1$, 
a standard pre-big bang model \cite{homepage, lidsey, sfd, pbb}
corresponds to $A < 0, B=-1$. 
Integration constants $a_{0}, \phi_{0}$ are equal to 
values of $a(t), \phi(t)$ at an initial moment $t_{i}= \frac{1-t_{0}}{A} $    
and $A$ determines initial values for $\dot a(t)$, $\dot \phi(t)$.
    
An one-to-one correspondence between post-big bang solutions and
pre-big bang solutions can be arranged by  duality  transformations
\beq \label{duality1}
a(t) \rightarrow \bar a(t) = a^{-1}(-t), \qquad 
\phi(t) \rightarrow \bar\phi(t) = \phi(-t) - 6 \ln a(-t).
\eeq
In terms of general solution (\ref{dg2}), (\ref{dg1}), 
the correspondence between a particular post-big bang solution ($B=1$) with 
$ a_0, \phi_0$  and a pre-big
bang solution ($B=-1$) is encoded in integration constants 
$\bar a_0, \bar\phi_0, A $
that have undergone the same duality transformation (\ref{duality1}):   
\be \label{predg2}
\bar a(t) = a_{0}^{-1} (t_{0} - At )^{-\frac{1}{\sqrt{3}}}\,, 
\ee
\be \label{predg1}
\bar\phi(t) = \phi_{0} - 6 \ln a_{0} - ({\sqrt{3} + 1 }) \ln (t_{0} -At). 
\ee

\subsection{Dilaton Gravity and Perfect Fluid Matter}

 After some manipulations the field equations become
\be   \label{doth}
\dot{H} + 3H^2 - H \dot{\phi} = \frac{1}{2} e^\phi \rho (\Gamma -1)\,, 
\ee
\be  \label{seos}
6H^2 -6 H {\dot\phi} + {\dot\phi}^2 = e^{\phi} \rho \,,
\ee
\be  \label{dotphi}
\ddot\phi + 3 H {\dot\phi} - {\dot\phi}^2 = 
    \frac{1}{2} e^\phi \rho (3 \Gamma - 4) \,,
\ee
\be  \label{pidev}
\dot\rho + 3 H \Gamma \rho = 0 \,, 
\ee
where the last equation is the  conservation law of matter. 
Loosely speaking, equations
(\ref{doth}), (\ref{dotphi}) and (\ref{pidev}) describe the evolution of 
$H$, $\phi$ and $\rho$, remaining equation (\ref{seos}) imposes
a constraint. 

A quasi-isotropic  solution of the field equations 
can be written as follows
\beq \label{postbb1}
 a &=& a_{0} (t_{0} \pm |\chi| t)^
         {\frac{2(\Gamma-1)}{\Delta}} \,,
\\ \label{postbb2}
 \phi &=& \phi_{0} - 
        \frac{2(4-3\Gamma)}{ \Delta } 
         \ln (t_{0} \pm |\chi| t) \,,  
\\ \label{postbb3}
 \rho &=& \rho_{0} (t_{0} \pm |\chi| t)^
        {-\frac{6 \Gamma(\Gamma-1)}{ \Delta}}\,,
\eeq
where 
\beq \label{hii}
  \chi \equiv \pm {\sqrt{\frac{e^{\phi_{0}} \rho_{0} \ 
         (\Delta)^2}{-4 \ (\Delta - 2)}}} \,,
\eeq
\beq \label{delta}
\Delta \equiv 3 (\Gamma -1)^2 +1  >0 \,,
\eeq
and $a_{0}$, $\phi_{0}$, $\rho_{0}$ are  integration constants which
are equal to the values of functions 
$a(t), \phi(t), \rho(t)$ at an initial moment 
$t_{i} = \frac{1-t_0}{\pm |\chi|}$. 
The solution is not the most general one, since initial values of
$\dot a(t)$ and $\dot \phi (t)$ are both uniquely determined by
barotropic index $\Gamma$ and the initial values of  
$\phi(t)$ and $\rho(t)$; in the most general case they must contain
an additional arbitrary integration constant. As a result,
solution (\ref{postbb1})--(\ref{delta})
turns out to be singular at $\Gamma_s = 1 \pm \frac{1}{\sqrt{3}}$, where
$\Delta -2 = 0$, $\chi \rightarrow \infty$. Note that 
in the case of unusal matter
with $(\Gamma -1)^2 > 1/3$ we must take $\rho_0 < 0$.
For the zero order three-velocity $u_i$, field equation (\ref{12}) implies 
$u_{i} \equiv 0$.
The range of variation of the time coordinate 
$t$ must be chosen such that $t_0 \pm |\chi| t \geq 0$; $t_0$ is a
numerical constant corresponding to the freedom of constant shift in time.
In comparison with solutions of the pure dilaton gravity
(\ref{dg2})--(\ref{dg1}), 
the integration constant
$A$ is replaced by constant $\chi = \chi (\rho_0, \phi_0, \Gamma)$ and   
the choice of the branch, $B = \pm 1$, is imitated by the choice of
barotropic index $\Gamma$, or more exactly, by the sign of expression
$\Gamma - 1 \in [-1, +1]$. 

Solution (\ref{postbb1}) for $a(t)$ determines 
the Hubble parameter as  
\be \label{H}
H=  \frac{\pm|\chi|}{(t_0 \pm |\chi|t)}\, \frac{2}{\Delta}\, (\Gamma -1),
\qquad
\dot H= \frac{-\chi^{2}}{(t_0 \pm |\chi|t)^{2}}\, \frac{2}{\Delta}\, 
        (\Gamma -1).  
\ee
We see that constant $\chi$ is proportional 
to the initial value of the Hubble
parameter at the initial moment $t_i$, $t_0 \pm |\chi|t_i =1$, and
in the case of matter with barotropic index 
$\Gamma_s = 1 \pm \frac{1}{\sqrt{3}}$ it diverges. In what follows we
don't consider barotropic indices belonging
to the neighborhood of $\Gamma_s$.

If we take $\Gamma >1$
and temporal argument $(t_0+ |\chi|t)$ we get a 
decelerating post-big bang Universe, e.g. for 
$\Gamma= \frac{4}{3}$ (radiation dominated stage) we have  
$|\chi|= \sqrt{{\frac{2}{3}} e^{\phi_0} \rho_0}$ and
$a(t) = a_{0}(t_{0} + |\chi|t)^{\frac{1}{2}}$,
matter density is decreasing,  $\rho = \rho_{0} (t_{0} + |\chi| t)^{-2}$, 
and  dilaton is freezed to a constant, $\phi = \phi_{0}$.
Integration constants 
$a_{0}$ and $\rho_{0}$    are 
(initial) values of corresponding variables at the moment 
$t_i = \frac{1 -t_{0}}{|\chi|}$.
A singularity ($a \rightarrow 0$,   $\rho 
\rightarrow \infty$) is reached at the moment 
$ t_{s} = -\frac{t_{0}}{|\chi|}$. 
The domain of the time variable is $t \in (t_{s},+\infty)$.
Taking into account our choice of  initial moment $t_i$ we see 
that $t_i = t_s + \frac{1}{|\chi|}$, i.e. $\frac{1}{|\chi|}$ is the
time interval from the singularity to the initial moment. If 
$t_0 >0$ we have  $t_{s} \leq t_{i} \leq 0$ 
and the Universe is regular  at the moment $t=0$.
If the source matter is exotic ($\Gamma < 1$), a solution 
(\ref{postbb1})--(\ref{postbb3}) represents a contracting Universe
beginning from a singularity at $t_s$ and going through $t_i$. 

The pre-big bang branch of solutions corresponds to the minus sign 
in front of $|\chi| t$  and exotic matter  $\Gamma -1 < 0$. 
The domain of the time coordinate is  
 $ t\in (-\infty, \frac{t_0 }{|\chi|} \equiv t_\infty)$, 
i.e. $t_0 - |\chi|t>0$, 
and    the behaviour of the scale factor is superinflationary
($\dot a > 0$, $\ddot a > 0$, $H > 0$, $\dot H > 0$).
Integration constants
$a_{0}$, $\phi_{0}$, and $\rho_{0}$
are the values of corresponding variables at the moment 
$ t_{i} \equiv \frac{1- t_0}{-|\chi|}$ and here 
$t_i = t_s -\frac{1}{|\chi|}$, i.e. $\frac{1}{|\chi|}$ is the time 
interval from the initial moment to the singularity.
A contracting solution with a future singularity is obtained 
if $\Gamma > 1$. 

Expressions (\ref{postbb2}) for $\phi$ and  (\ref{H}) for $H$   imply  
\be
  H = - (\Gamma -1)\dot{\bar\phi}, \qquad
            \dot{\bar\phi} \equiv \dot{\phi}  - 3H
\ee
and the  behaviour of solutions depending on barotropic index 
$\Gamma$
can be summarized on a phase diagram analogous to the corresponding one
familiar from  the pre-big bang scenario of the pure dilaton gravity:

\begin{figure}[h]
\centering
\vskip -3cm
\hskip 0.5cm \psfig{file=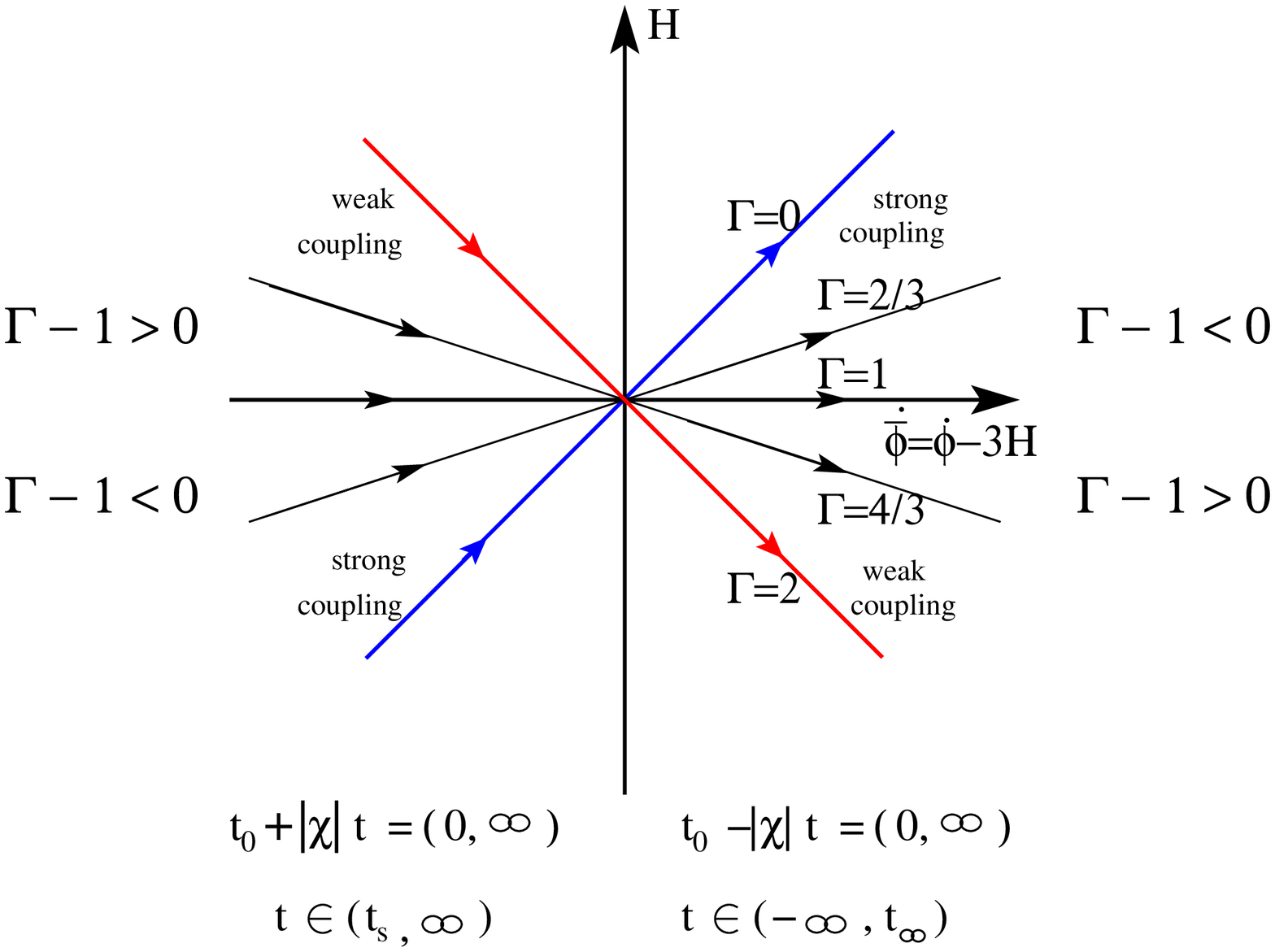,width=11cm}
\vskip -3.5cm
\label{figure1}
\end{figure}

As in the case of pure dilaton gravity,  solution 
(\ref{postbb1})--(\ref{postbb3}) contains both the post-big bang
and the  pre-big bang branch. The latter ones  can 
be related to the former ones by  
duality transformations (\ref{duality1}), 
but in order to satisfy the field equations, here  we need an 
additional transformation
for barotropic index $\Gamma$ and for energy density $\rho$:
\beq \label{duality2}
{\frac{p}{\rho}} \rightarrow {\frac{\bar p}{\bar\rho}} = -{\frac{p}{\rho}}
 ~~~ \Longrightarrow {\bar \Gamma} \rightarrow 2 -\Gamma\,,
\qquad \rho  \rightarrow \rho a^6(-t)\, . 
\eeq 
The corresponding pre-big bang solutions (i.e. with the time dependence
in the form $ ( t_{0} - | \chi| t)$)   now read 
\beq \label{prebb1}
 {\bar a} &=& a_{0}^{-1}\, ( t_{0} - | \chi| t)^
          {\frac{2({\bar \Gamma}-1)}{ \Delta }} \,,
\\   \label{prebb2}
 {\bar\phi} &=& \phi_{0} - 6 \ln a_{0}  - \ 
        \frac{2(4 - 3\bar\Gamma)}{ \Delta} \ 
        \ln ( t_{0} - | \chi| t) \,, 
\\   \label{prebb3}
 {\bar\rho} &=& \rho_{0} a_0^6 \, ( t_{0} - |\chi| t)^
        {-\frac{6 \bar\Gamma(\bar\Gamma-1)}{\Delta}}\,.
\eeq
Note that $\rho_0 e^{\phi_0} = {\bar{\rho}}_0 e^{\bar{\phi}_0}$,
 $\Delta = \bar\Delta $ and $t_{0}$ don't change under  duality
transformations (\ref{duality1}), (\ref{duality2}).

\subsection{\textbf{Model building}}

In the case of pure dilaton gravity, the main idea of the original 
pre-big bang cosmology was to use
a superinflationary pre-big bang branch obtained from a decelerating
post-big bang branch by duality transformations for constructing 
a cosmological model valid in the time range $t \in (-\infty, +\infty)$.
However, due to rather general no-go theorems \cite{kmo}   
a smooth connection of these two branches 
turned out to be impossible, at least without some kind of higher order 
corrections. 
In the case of a perfect barotropic fluid,
 mutually dual solutions (\ref{postbb1})--(\ref{postbb3}) and
(\ref{prebb1})--(\ref{prebb3}) contain different   
 barotropic indices $\Gamma, \bar \Gamma = 2 - \Gamma$  and consequently 
also different matter sources. 
This means that if we try to repeat the strategy
of the original pre-big bang cosmology, we must admit a profound   
change in the type of cosmological matter and 
join a pre-big bang solution 
with barotropic index $(2- \Gamma)$ with a solution  
with barotropic index $\Gamma$ (cf \cite{ellis, is}). 
We are free to choose the moment of branch change, 
but the explicit solution (\ref{H}) for $H$
indicates that we cannot achieve continuity for both $H$ and $\dot H$.

For example, 
let $a_0=1$ and $t_0=1$  and let the junction moment be the regular
point $t=0$. From the solutions (\ref{postbb1})--(\ref{postbb3}), 
(\ref{prebb1})--(\ref{prebb3}) it is easy to see that $a(t)$, $\dot a(t)$, 
$\phi (t)$ and $\rho(t)$ are continuous there, 
but not
$\dot H (t) $ (or $\ddot a(t)$),  $\dot \phi(t)$ and $\dot \rho (t)$:
\be
\dot {\bar H} (0)  = 2 \chi^2 (\Gamma -1) \Delta^{-1} =- \dot H(0) ,
\qquad \dot {\bar \phi} (0) = \dot \phi (0) -4 |\chi| \Delta^{-1}, 
\ee
\be
\dot {\bar \rho} (0) = \dot \rho (0) + 12 \rho_0 |\chi| (\Gamma -1)^2
\Delta^{-1}.
\ee 
It is possible, that these kinematical discontinuities 
could be given in terms of physical changes of matter sources. 
However, a dynamical description of these changes remains open.  
Numerical values 
of discontinuities depend on the numerical value of the constant $\chi$
given in terms of barotropic index $\Gamma$ and 
initial values of the dilaton and matter density
(\ref{hii}).  We can choose $\frac{1}{|\chi|} \sim L_{\rm Planck} $;
then  discontinuities are nearly vanishing, but
initial moments are almost in the singularities and higher order
corrections must be taken into account. Alternatively, we can  choose 
initial moments to be in a classically regular region, 
far from the singularities; then, however, discontinuities
are finite.  

The other possibility is 
to abandon duality transformations altogether and to connect 
solutions with the same barotropic index $\Gamma$, but with different
temporal arguments $(t_0 \pm |\chi|t)$ and  with 
arbitrary integration  
constants $a_{0}$, $\phi_{0}$, $\rho_{0}$, $\chi$  and $t_0$
(cf ekpyrotic models \cite{kosst, ekpyrotic}).  However, according 
to the solution (\ref{H}) if 
$\dot H$ is continuous, $H$ is not.

As already mentioned, the moment of branch change can be
chosen arbitrarily and  it may be a regular or a singular point of
the solution. This means that the problem of graceful exit is
distinct from the problem of singularity. The graceful exit
problem in the string cosmology is in fact very similar to that
of ordinary inflationary cosmology, where accelerating inflationary stage 
must be joined with a decelerating FRW model. In ekpyrotic models, 
discontinuities  are concealed in the singularity.


\section {Second Order Solutions} \label{neli}

Let us we adopt the procedure presented in \cite{cdlp, cdl}
to equations (\ref{11})--(\ref{14}) derived from the string 
effective action (\ref{3+1}).
The second order solutions for the three-metric, dilaton,
energy density and three-velocity are assumed to be in the form  
\beq \label{2j1}
 \gamma_{ij}(t,x^{k}) = a^{2}(t) \left[ h_{ij} + f_{2}(t) R_{ij}(h_{kl}) + 
            {\frac{1}{3}}(g_{2}(t) - f_{2}(t)) R(h_{kl}) h_{ij} \right] ,
\eeq
\beq \label{2j2}
 \phi(t,x^{k}) &=& \phi(t) + \phi_{2}(t) R(h_{kl}) ,\\
 \rho(t,x^{k}) &=& \rho(t) + \rho_{2}(t) R(h_{kl}) ,\\
  u_{i}(t,x^{k}) &=& u(t) \nabla_{i} R(h_{kl}) .
\eeq
Here, $R_{ij}(h_{kl})$ and $R(h_{kl})$ are the 3-dimensional  
Ricci tensor and the scalar 
curvature formed from the seed metric $h_{kl}$, respectively.
The zero order solutions $a(t)$, $\phi(t)$, and $\rho(t)$ 
are given in the previous section.
Substituting them into field equations, keeping only terms up 
to the second order in spatial gradients, collecting  coefficients at
$R^{j}_{i}$, and $R$ we get a system of ordinary differential equations 
for the second order corrections $f_{2}$, $g_{2}$, and $\phi_{2}$.

\subsection{Pure Dilaton Gravity}

If we substitute  expressions (\ref{2j1})--(\ref{2j2}) 
into field equations (\ref{11})--(\ref{14}) and take $\rho = 0$ we get 
equations for the second order corrections 
$f_{2}(t)$, $g_{2}(t)$, and $\phi_{2}(t)$ 
\beq \label{dg3}
  \ddot f_{2} + (3 H - \dot \phi) \dot f_{2} = -2 a^{-2} , 
\eeq
\beq \label{dg4}
 (2H - \dot \phi) \dot g_{2} - (6H - 2 \dot \phi) \dot \phi_{2} = - a^{-2},
\eeq
\beq \label{dg5}
 \ddot \phi_{2} + (3H - 2 \dot\phi) \dot \phi_{2} 
      + \frac{1}{2} \dot \phi \dot g_{2} = 0.
\eeq
Upon substituting zero order solutions (\ref{postbb1})--(\ref{postbb3})
into equations (\ref{dg3})--(\ref{dg5}) we get the second order 
solutions (we have taken $A=1$)    
\beq \label{dg6}
f^\pm_{2}(t) &=& c^\pm_{1} + c^\pm_{2} \ln (t_{0} \pm t) - 
{\frac{3 (\sqrt{3} \pm 1)}{4 (\sqrt{3} \mp 1)}} 
(t_{0} \pm t)^{2 \mp \frac{2}{\sqrt{3}}} ,
\\[2ex]
\phi^\pm_{2}(t) &=& e^\pm_{1} + {\frac{\sqrt{3} \mp 1}{2 \sqrt{3}}}d^\pm_{2} 
(t_{0} \pm t)^{-1} + 
{\frac{3 \sqrt{3}}{4 (11 \mp 5 \sqrt{3})}} 
(t_{0} \pm t)^{2 \mp \frac{2}{\sqrt{3}}} \label{dg7} ,
\\[2ex]
g^\pm_{2}(t) &=& d^{\pm}_{1} + d^\pm_{2} (t_{0} \pm t)^{-1} - 
{\frac{3 (4 \sqrt{3} \mp 7)}{2 (21 \sqrt{3} \mp 37)}} 
(t_{0} \pm t)^{2 \mp \frac{2}{\sqrt{3}}} \label{dg8} .
\eeq
 The first  terms in expressions 
(\ref{dg6})--(\ref{dg8}) are
solutions of corresponding  homogeneous equations  and
the last ones are particular solutions of inhomogeneous equations.

As indicated by Comer, Deruelle and Langlois \cite{cdl} terms
proportional to $(t_{0} \pm t)^{-1}$ which diverge at 
$(t_{0} \pm t) \rightarrow 0$   
can be removed by 
an infinitesimal coordinate transformation that preserves the synchronous
coordinate system.  
In the second approximation, the transformation of the three-metric  
can be given as
$\gamma^{'}_{ij} = \gamma_{ij} - 2 a \dot{a} T_{2} R(h_{ij}) h_{ij}$
and the transformation of the dilaton as $\phi^{'}= \phi - \dot\phi T_{2} R$ 
with $T_{2}= const$ \cite{cdl}.  
If we take $T_{2} = \frac{d^{\pm}_{2}}{2 \sqrt{3} }$, then terms
proportional to $ d_2^\pm (t_{0} \pm t)^{-1}$ cancel.
The logarithmic  term in $f_{2}$ 
cannot be removed by a coordinate transformation (an analogous
problem occurs in the case of linear perturbation theory \cite{vmuh}), but
probable it can be treated by means of a renormalization procedure \cite{nambu}.

In the case of  lower signs (pre-big bang)  solution 
(\ref{dg6})--(\ref{dg8}) coincides with the solution of the
corresponding Hamilton-Jacobi equation in the second approximation  
 presented by us
earlier  and used for estimating 
the size of an initial homogeneous domain
for getting enough inflation in the pre-big bang stage \cite{kuusk1}.


\subsection {Dilaton gravity and perfect fluid matter}

Following the procedure described above and eliminating the matter 
density $\rho$ from field equations 
we get a system of equations for $f_{2}(t)$, 
$g_{2}(t)$, and $\phi_{2}(t)$ 
\beq \label{pf1}
  \ddot f_{2} + (3 H - \dot \phi) \dot f_{2} = -2 a^{-2},
\eeq
\beq \label{pf2}
  \ddot g_{2} + \left[ 6 H - \frac{(7-6\Gamma)}{(4-3\Gamma)} \dot \phi \right] 
  \dot g_{2} &-&
  \left[\frac{6(7-6\Gamma)}{4-3\Gamma} H - 
        \frac{12(1-\Gamma)}{(4-3\Gamma)} \dot \phi \right] \dot \phi_{2} 
\nonumber\\[1ex]
   &-& \frac{6(1-\Gamma)}{4-3\Gamma} \ddot \phi_{2} = -2 a^{-2} ,
\\[2ex]
\label{pf3}
   \ddot g_{2} + \left[ 2 H - \frac{1}{(4-3\Gamma)} \dot \phi \right] \dot g_{2} &-&
   \left[\frac{6}{4-3\Gamma} H - \frac{4}{(4-3\Gamma)} \dot \phi \right] \dot \phi_{2}
\nonumber\\[2ex] 
    &-& \frac{(10-6\Gamma)}{4-3\Gamma} \ddot \phi_{2} = 0.
\eeq
These equations are apparently singular at $\Gamma = \frac{4}{3}$ but 
in fact they are not because of the form of the solution in the zero 
approximation.

Let us substitute the first order solutions 
(\ref{postbb1})--(\ref{postbb3}) into  equations (\ref{pf1})--(\ref{pf3}).
Upon long but straightforward calculations we obtain the following
system of equations
\beq \label{pf4}
 \ddot f^\pm_{2} &\pm& \frac{2}{\Delta} |\chi| 
    (t_{0} \pm |\chi|t)^{-1} \dot f^\pm_{2} = - 2 a_{0}^{-2} (t_{0} \pm|\chi|t)^
    {-\frac{4(\Gamma-1)}{\Delta}},\\[2ex]
\label{pf5}
 \ddot g^\pm_{2} &\pm& 2 |\chi| (t_{0} \pm |\chi| t)^{-1} \dot g^\pm_{2} = 
     (3 \Gamma -5) a_{0}^{-2}(t_{0} \pm|\chi|t)^
    {-\frac{4(\Gamma-1)}{\Delta}},\\[2ex]
\label{pf6}
 \ddot \phi^\pm_{2} &\pm& \frac{2}{\Delta} 
                |\chi| (t_{0} \pm |\chi|t)^{-1} \dot \phi^\pm_{2}
                 - \frac{4-3\Gamma}{10-6\Gamma} \ddot g^\pm_{2}
\nonumber\\[1ex] 
              &\mp& \frac{(2\Gamma-1)(4-3\Gamma)}{(5-3\Gamma) \Delta}
                |\chi| (t_{0} \pm |\chi|t)^{-1} \dot g^\pm_{2} = 0.
\eeq
The upper sign corresponds to the post-big bang behaviour and the lower sign
corresponds to the pre-big bang 
behaviour.
Retaining all integration constants  general solutions of equations 
(\ref{pf4})--(\ref{pf6}) can be written
\beq \label{f2}
f^\pm_{2}(t) &=& c^\pm_{1} + c^\pm_{2}(t_{0} \pm |\chi| t)^
         {\frac{3\Gamma^2 - 6\Gamma + 2} \Delta}
\nonumber\\[1ex]        
        &-&{\frac{ \Delta^2 |\chi|^{-2} a_{0}^{-2}} 
        {(3\Gamma^2 - 8\Gamma + 6)(3\Gamma^2 - 10\Gamma + 10)}}
        (t_{0} \pm |\chi| t)^{2 - \frac{4(\Gamma - 1)} {\Delta}},\\[2ex]
g^\pm_{2}(t) &=& d^\pm_{1} + d^\pm_{2}(t_{0} \pm |\chi| t)^{-1} 
\nonumber\\[1ex]  
         & +&\frac{(6\Gamma-10) \Delta^2 |\chi|^{-2} a_{0}^{-2}}
        {4(3\Gamma^2 - 8\Gamma + 6)(9\Gamma^2 - 22\Gamma + 16)}
        (t_{0} \pm |\chi| t)^{2 - \frac{4(\Gamma - 1)} {\Delta}} \label{g2},\\[2ex]
\phi^\pm_{2}(t) &=& e^\pm_{1} + e^\pm_{2}(t_{0} \pm |\chi| t) 
            ^\frac{3\Gamma^2 - 6\Gamma + 2}{ \Delta} 
            + \frac{3\Gamma - 4}{6(\Gamma - 1)} d_{2}^{\pm} (t_{0} 
              \pm |\chi| t)^{-1}
\nonumber
\eeq
\beq \label{fii2}
  + \frac{(3 \Gamma -4)(3\Gamma^2 - 6\Gamma + 6) \Delta^{2} 
           |\chi|^{-2} a_{0}^{-2}} 
 {4(3\Gamma^2 - 8\Gamma + 6)(9\Gamma^2 - 22\Gamma + 16)
 (3\Gamma^2 - 10\Gamma + 10)}
 (t_{0} \pm |\chi| t)^{2 - \frac{4(\Gamma - 1)} {\Delta}}. 
\eeq
For the matter density correction $\rho_{2}$ we get from equation (\ref{14})
\beq
  \rho_{2} = \frac{2}{(3\Gamma-4)} ~ e^{-\phi} \left[ \ddot \phi_{2} 
              + 3 H \dot \phi_{2}
             + \frac{1}{2} \dot g_{2} \dot \phi + 2 \dot \phi \dot \phi_{2}
             - \frac{1}{2} (3\Gamma-4) e^{\phi} \rho \phi_{2} \right]
\eeq
and for the three-velocity we get from equation (\ref{12})
\beq
  u(t) = \frac{2 e^{-\phi}}{ \Gamma \rho } \left[\frac{1}{3} \dot g_{2} - \frac{1}{12} 
  \dot f_{2} - H \phi_{2} - \dot \phi_{2} \right]. 
\eeq
For the density contrast we can write the expression as follows
\beq
   \delta &=& \frac{\rho_{2}}{\rho} = \frac{2}{(3\Gamma-4)\rho_{0} e^{\phi_{0}}} 
            (t_{0} \pm |\chi| t)^{2} \times 
\nonumber\\[1ex]
          & &\left[ \ddot \phi_{2} + 3 H \dot \phi_{2}             
            + \frac{1}{2} \dot g_{2} \dot \phi + 2 \dot \phi \dot \phi_{2}
             - \frac{1}{2} (3\Gamma-4) e^{\phi} 
             \rho \phi_{2} \right] .
\eeq

Let us consider the behaviour of  inhomogeneities 
in the pre-big bang case near the final singularity  
 ($|\chi|t \rightarrow t_{0}$).  
 Terms proportional to 
$(t_{0} - |\chi|t)^{\frac{3\Gamma^2 - 6\Gamma + 2}{\Delta}}$ are 
decaying if 
$\Gamma  < 1-\frac{1}{\sqrt{3}} $ and $\Gamma > 1+\frac{1}{\sqrt{3}}$,
e.g. at $\Gamma =0$  and $\Gamma = 2$. 
Terms proportional
$(t_{0} - |\chi|t)^{2 - \frac{4(\Gamma - 1)} {\Delta}}$ 
are decaying at all values of $\Gamma$.  
Terms proportional to
$(t_{0} - |\chi|t)^{-1}$ can be treated as in  
the case of pure dilaton gravity:
they can be cancelled by an analogous coordinate transformation with 
$T_{2}= \pm \frac{\Delta d^{\pm}_{2}} {12 (\Gamma-1) |\chi|}$.

In the post-big bang case, the model with  
$\Gamma = \frac{4}{3}$ should describe
a radiation dominated FRW cosmology and so it does in the zero
approximation. 
However, in the second approximation the
initial inhomogeneities are not decaying at late times $t \gg t_0$,
although logarithmically divergent terms that occur in the
solution for the pure dilaton gravity are absent here. 
The full expression for the metric (\ref{2j1}) includes  terms  
proportional to  $(t_{0} + |\chi|t)^2$ which are growing at all
values of $\Gamma$. An analogous term also appears 
in solutions of field equations corresponding to  other  
models  \cite{cdlp}, \cite{cdl}, \cite{lks}. 
Khalatnikov, Kamenshchik, and Starobinsky \cite{lks}  
have recently   indicated 
its inevitability near the singularity. However, 
in our solution  and in the solutions presented in  \cite{cdlp} 
it also appears  in the late time evolution.



\section {Summary} 

In this paper we investigated  cosmological solutions of 
equations (5)--(8), derived from the low energy effective string action,
in the framework of long-wavelength approximation.
The zero  order solution (\ref{postbb1})--(\ref{delta}) contains both the 
pre-big bang and the post-big bang branch, but as it is well known, it is not 
possible to connect these two branches smoothly.
We have found, that the problem of singularity and the problem of
graceful exit may be disconnected. The graceful exit problem resembles 
the problem  already known from the first models of inflation: 
how to get a smooth
transition from an accelerating background to a decelerating background.
The transition moment and the moment when classical solutions become singular
may be different, in general. 
On the other hand, as indicated in the context of ekpyrotic scenario the 
occurrence of singularity is strictly a four-dimensional phenomenon and may be
used to hide a nonsmooth transition.  

The second order corrections to the long-wavelength solutions include the 
effect of spatial gradients. 
As expected, the picture is symmetric: terms which are growing in the 
pre-big bang phase are decaying in the post-big bang phase and vice versa;
we might use different terms in different asymptotic
regimes to estimate the significance of corrective terms.  
We conclude that in the case of an exotic matter 
with barotropic index $\Gamma = 0$ or $\Gamma =2$ 
we can achieve the decay of all second order correction terms
at the end of the pre-big bang stage. However, corrections for
the post-big bang stage always include terms which are growing in time. 
Our conclusion indicates that a simple model without any dilaton potential 
or cosmological constant leads to a result which can hardly fit  our 
present understandings.
Inclusion of additional terms in the field equations
(e.g. $V(\phi) $, $\Lambda$, or $H_{\mu\nu \rho}$)   changes
the form of solutions significally and in that case  the analysis 
of the second order is far more complicated.

\bigskip

\textbf{ACKNOWLEDGEMENTS}
\smallskip

This work was supported by the Estonian Science Foundation 
under grant No 5026.

\end{document}